\documentclass[aps,prb,reprint,longbibliography]{revtex4-2}

\usepackage{graphicx}  
\usepackage{subfigure}
\usepackage{multirow}

\usepackage{longtable}
\usepackage{parskip}
\usepackage[T1]{fontenc}
\usepackage{dcolumn}   

\usepackage{bm}        
\usepackage{amsfonts}  
\usepackage{amsmath}   
\usepackage{amssymb}   
\usepackage{placeins}

\begin{document}

\title{Nuclear-Electron Hyperfine Coupling of the Shallow States Associated with Vacancies in Gallium Nitride}

\author{Joseph Sink$^1$}
\author{Michael E. Flatt\'e$^{1,2}$}

\affiliation {\it $~^{1}$Department of Physics and Astronomy, University of Iowa, Iowa City, Iowa 52242, USA\\
$~^{2}$ Department of Applied Physics and Science Education, Eindhoven University of Technology, Eindhoven, Netherlands}

\date{\today}

\begin{abstract}  
We use multiband real space Green's functions computed using open-boundary conditions for clean GaN to exactly solve the potential-scattering Dyson equation to obtain the electronic structure of single 
nitrogen and gallium vacancies. From these vacancy solutions, we compute the local density of states as well as the Fermi contact and anisotropic contributions to the hyperfine field in the vicinity of the defect. These quantities directly affect electrically-detected magnetic resonance signals, which can be used to identify these defects when present in GaN devices. 
\end{abstract}

\maketitle 

\section{Introduction}

Characterizing and identifying defects in semiconductors is an essential first step for developing strategies to reduce their prevalence in and/or impacts on device performance. 
Physical models of the spin-coherent dynamics of defects combined with magnetic resonance measurements has proven to be an indispensable method to achieve this goal. Such techniques include electron spin/paramagnetic resonance (ESR/EPR) and electrically/optically detected magnetic resonance (EDMR/ODMR) \cite{Lenahan.JAP.1984,kim.JAP.1988,Nestle.PRB.1997}. In these experiments, the defect electronic spin $g$-tensor and the hyperfine interactions  $\overset{\leftrightarrow}{A}_i$ from all nuclei $\{i\}$ largely determine the resonance frequencies and line shapes.
Magnetic resonance has been used extensively to characterize group-IV materials (i.e., C, Si and Ge) which have a naturally low abundance of spinful nuclei. Defects in these systems may be coupled to few if any spinful nuclei (e.g., the NV center in diamond\cite{Fortman.JPCA.2019,Jelezko.PSS.2006}). The weak coupling to the nuclear spin bath in these materials 
 yields long spin-coherence lifetimes, sharp resonances, and an ability to test map the relationship between defects and their local nuclear-spin environment\cite{Zhao.NatNano.2012}.

GaN has emerged as a leading candidate for next-generation electronic and optoelectronic devices because of its wideband gap, high thermal conductivity, and robustness under extreme conditions. Despite the technological importance of GaN, the microscopic identities of several electrically active defects remain undetermined\cite{Carlos.PRB.1993,vonBardeleben.PRL.2012}. 
However, in contrast to the group-IV materials, all (stable) group-III and -V nuclei are spinful ($I\neq0$) with III-V materials having characteristically short spin-coherence lifetimes. This makes identifying defects in III-V's using conventional group-IV EPR and EDMR challenging.

To address some of these limitations, a variant of the EDMR technique, operated at near-zero applied DC magnetic field (NZFMR)\cite{Myers.JAP.2022,Frantz.APL.2021,Moxim.JAP.2024}, may prove useful. NZFMR is an all-electrical method that measures the differential magnetoresistance under static DC bias as the magnetic field is adiabatically swept through zero. The measured magnetoresistance arises from a two defect state Pauli spin blockade, where the system becomes trapped in a triplet spin configuration and is relieved when it transitions to a singlet configuration. The rate of this transition depends on the relative precession rates of the two electron spins, which at zero field are dictated by the local hyperfine interactions.

{Native vacancies in GaN (V$_\mathrm{N}$ and V$_\mathrm{Ga}$) exhibit shallow donor- or acceptor-like behavior for the charge neutral charge state\cite{Jenkins.PRB.1989,VandeWalle.JAP.2004,Lyons.npjCompMat.2017} which are associated with extended hydrogenic bound-state envelopes.
Such delocalization poses a well-known challenge for supercell-based {\it ab initio} approaches, as the defect wave function and its long-range Coulomb potential overlap significantly with their periodic images, leading to slow convergence with supercell size\cite{VandeWalle.JAP.2004,freysoldt.RevModPhys.2014,Komsa.APS.2012}.
While deep, strongly localized defect states may converge with relatively modest cells, reliably describing shallow states requires supercells with dimensions multiple times larger than the effective Bohr ($a^*$) radius\cite{Swift.npjCompMat.2020}, and consequently a computational cost that grows rapidly with the required field of view (roughly with the supercell volume)

To circumvent these limitations, we utilize a real-space tight-binding Green's-function framework to solve for a single non-interacting localized potential. 
The Green's functions are computed from a well established empirical tight binding model \cite{Jancu.APL.2002} for GaN that gives an adequate representation of the gap, valence band splittings and gap extrema effective masses. 
From the resulting inhomogeneous (i.e., dressed) Green's function solutions, we extract the wave function and hyperfine couplings for nearby nuclei, enabling future NZFMR and EDMR simulations of these defects.

This article continues with (Section \ref{sec:methods})  an introduction to the methods and formalism pertaining to the calculation of the Green's functions in the presence of a signle defect in an otherwise infinite clean host, along with extraction of several observable quantities. Tabulated results for the hyperfine fields for near-neighbor nuclei and discussion are listed in Section \ref{sec:results}. A brief concluding section is provided.

\begin{figure*}[ht!]
    \centering
    \includegraphics[width=1\linewidth]{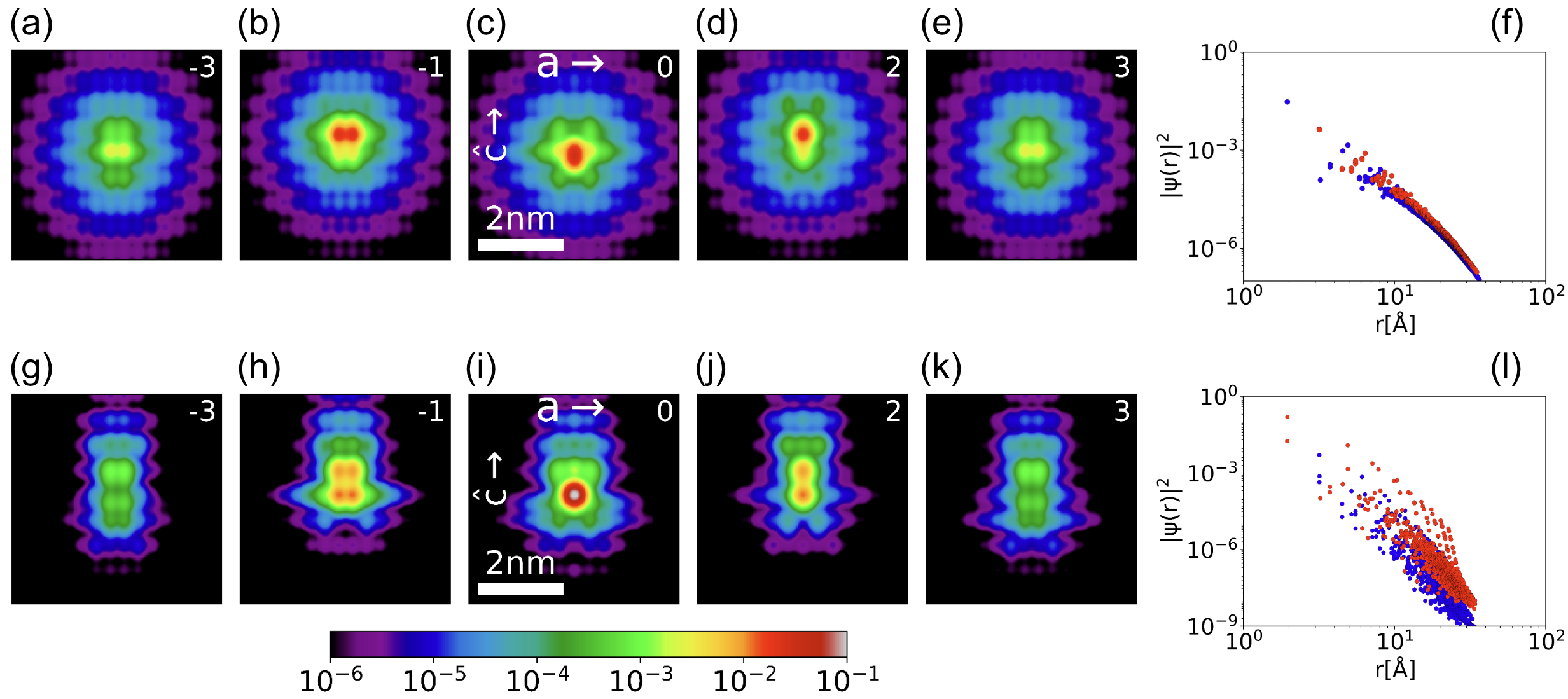}
    \caption{$|\psi(\mathbf{r})|^2$ at various distances along the $\hat{m}$ plane for the $V_{N}$(a-e) and $V_{Ga}$(g-k) defect. The depth along $\hat{m}$ measured from the layer containing the defect is shown in the top right of each panel in (a-e) and (g-k) in steps of 0.92059 \AA. Staggered layer depth is due to the way the $\hat{m}$-plane intersects the lattice. The corresponding radial decay of the defect wave function for the $V_{N}$ and $V_{Ga}$ defects are shown (f) and (l), respectively. The nitrogen shallow donor displays a delocalized spherical distribution. By contrast, the gallium defect is highly localized and anisotropic. }
    \label{fig:wavefunction}
\end{figure*}
\section{Methods}\label{sec:methods}

\subsection{Defect Calculation}
We treat the single non-interacting defect problem using a Green's function scattering method. This method has been used successfully to calculate defect levels \cite{Hjalmarson.PRL.1980,Kobayashi.PRB.1983} and wave functions \cite{TangFlatte.PRL.2004,KortanFlatte.PRB.2017} in semiconducting systems. The defect properties are extracted from the solutions of the Dyson Equation,
\begin{align}
    \hat{G}=(1-\hat{V}'\hat{g})^{-1}\hat{g},\label{eq:Dyson_equation}
\end{align}
where $\hat{G}$ is the inhomogeneous solution for the system containing the defect, $\hat{g}$ is the homogeneous solution corresponding to the bulk system without the defect, and $\hat{V}'$ is the localized perturbative potential corresponding to the defect. Conceptually, this process can be visualized as bulk Bloch waves scattering off of a localized impurity potential.

The real-space Green's functions for the bulk system with open boundary conditions, $\hat{g}$, are computed by taking the inverse Fourier transform of the resolvent of the bulk k-space Hamiltonian, $\hat{H}_0(\mathbf{k})$,
\begin{align}
    &\hat{g}(\mathbf{R},\mathbf{R}';\omega)=\\ \label{eq:bulk_g}
    &\ \ \ \ \frac{V}{(4\pi)^3}\int _{\Omega_{BZ} } (\omega+\delta^+i-\hat{H}_0(\mathbf{k}))^{-1}e^{-i \mathbf{k}\cdot(\mathbf{R}-\mathbf{R}')}d^3k\nonumber
\end{align}
where $\omega$ is the spectral frequency, $\delta$ is the spectral linewidth, and $\mathbf{R}$ is an atomic site in the lattice. Note that these Green's functions are computed using open boundary conditions, and are free from both finite volume artifacts and periodicity related self-interactions. 
The integration is carried out numerically using an adaptive $4^{th}$ order Runga-Kutta Cash-Karp (RKCK) method\cite{RKCK} with an integration tolerance of $8\text{x}10^{-5}$. 

We use here a multiband spd$s^*$ with spin-orbit coupling tight binding (TB) model \cite{Jancu.APL.2002} which reproduces well the effective masses, splittings, and band gap for bulk wurtzite GaN.

\begin{figure}
    \centering
    \includegraphics[width=1\linewidth]{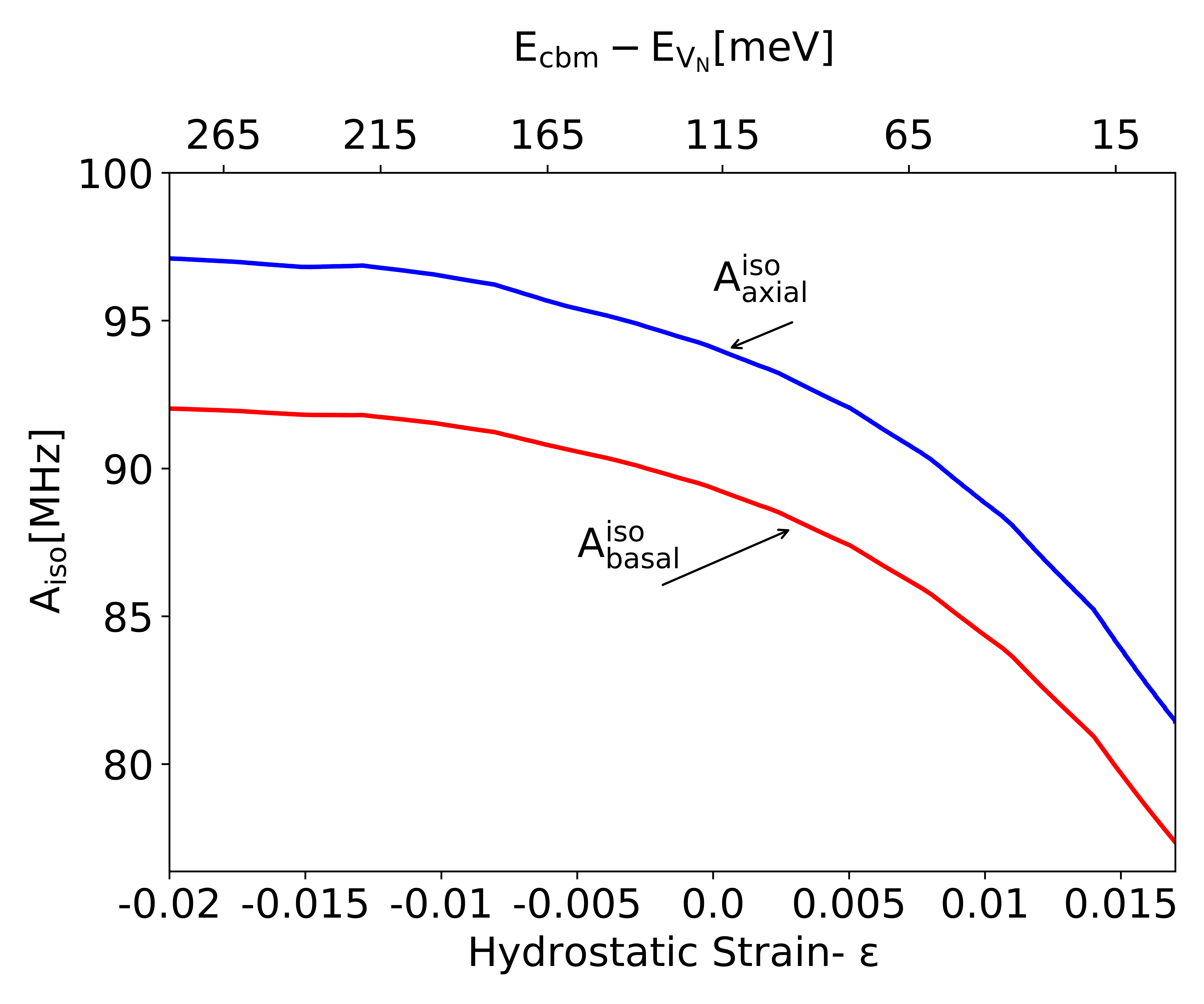}
    \caption{Nitrogen vacancy Axial and Basal hyperfine dependence on near neighbor induced strain. The corresponding bound state energy is shown on the top axis. As the bound state energy is decreased (more shallow donor state) the wave function effective radius increases resulting in a decrease in the isotropic Fermi contact hyperfine.}
    \label{fig:enter-label}
\end{figure}

For local perturbations, the Dyson Equation [Eq.~(\ref{eq:Dyson_equation})]
is exactly solvable and can be put in block diagonal form\cite{TangFlatte.PRL.2004,FlatteByers.PRB.1997,flattebyers.book}. These blocks correspond to the regions where the perturbation takes on non-zero or zero values (Fig. \ref{fig:near_far_cartoon}), and are referred to here as the `near-field' and `far-field', respectively. In block form, the perturbative potential and bulk Green's functions can be generically expressed as,
\begin{align}
    \hat{V}'=
    \begin{pmatrix}
        \hat{V}_{nn}&0\\
        0&0
    \end{pmatrix}&&
    \hat{g}=
    \begin{pmatrix}
        \hat{g}_{nn}&\hat{g}_{fn}\\
        \hat{g}_{nf}&\hat{g}_{ff}
    \end{pmatrix}.
\end{align}
Solving Eq. (\ref{eq:Dyson_equation}) with the above definition of the potential yields the inhomogeneous relations,
\begin{align}
    \hat{G}=
    \begin{pmatrix}
        \hat{M}_{nn}\hat{g}_{nn}&\hat{g}_{fn}\hat{M}_{nn}\\
        \hat{M}_{nn}\hat{g}_{nf}&~~~~\hat{g}_{ff}+\hat{g}_{fn}\hat{V}_{nn}\hat{M}_{nn}\hat{g}_{nf}
    \end{pmatrix}\label{eq:bulk_g_block}
\end{align}
where $M_{nn}=(\hat{1}-\hat{g}_{nn}\hat{V}_{nn})^{-1}$.

In the tight-binding basis, the impurity potential is,
\begin{align}
    \hat{V}'=\sum_{\beta,\beta'}V_{\beta,\beta'}|\beta\rangle \langle\beta'|
\end{align}
where $|\beta\rangle=|nl\alpha\chi;q\rangle$ is a composite index representing the atom site, $q$, atomic orbital $|nl\alpha\rangle$ (i.e., $\{ns,np_i,nd_{ij},etc.\}$) and spin $|\chi\rangle$.

\subsection{Observables}
The single particle inhomogeneous Green's function solutions are related to the density matrix and can be used to compute the expectation value of an observable, $\hat{O}$, via the relation,
\begin{align}
    \langle\hat{O}\rangle_\omega=\frac{-1}{\pi}Im\left[Tr\left[\hat{G}(\omega)\hat{O}\right]\right].\label{eq:Greens_function_expectation_value}
\end{align}
This relationship can be readily seen by expanding Eq. (\ref{eq:Greens_function_expectation_value}) in the spectral basis,
\begin{align}
    \langle\hat{O}\rangle_\omega=\frac{-1}{\pi}\lim_{\delta\rightarrow 0^+}Im\left[Tr\left[\sum_{n}\frac{|\psi_n\rangle\langle\psi_n|}{\omega+\delta i-E_n}\hat{O}\right]\right]\label{eq:op_spectral},
\end{align}
Which can be rewritten as,
\begin{align}
    \langle\hat{O}\rangle_\omega=\sum_{n}\langle\psi_n|\hat{O}|\psi_n\rangle \delta(\omega-E_n).
\end{align}
by taking advantage of the cyclic property of the trace and the Sokhotski–Plemelj theorem (i.e., $\frac{1}{\omega+0^{+}i}\rightarrow \frac{1}{\omega}\mathcal{P}-i\pi\delta(\omega)$).

Experimentally, the defect wave function can be measured via cross-sectional STM, which is sensitive to the local density of states, $\eta$. We compute $\eta$ by evaluating Eq. (\ref{eq:Greens_function_expectation_value}) with $\hat{O}=\langle\hat{R}\rangle$ (i.e., the position operator),
\begin{align}
    \langle\hat{R}\rangle_\omega=\nonumber\\
    =&\sum_{n}\langle\psi_n|\mathbf{R}\rangle\langle\mathbf{R}|\psi_n\rangle \delta(\omega-E_n)\nonumber\\
    \eta(\mathbf{R},\omega)=&\frac{-1}{\pi}Im[Tr[\hat{G}(\mathbf{R},\mathbf{R};\omega)]].\label{eq:LDOS}
\end{align}

The $|\psi|^2$ is computed by evaluating Eq. (\ref{eq:LDOS}) with $\omega$ resonant with a bound state of the system, $E_d$, for sites in both the near and far field. 

We take the isotropic (Fermi contact) and anisotropic dipole-dipole components of the hyperfine tensor within the point dipole approximation \cite{Blochl.PRB.2000} to be,
\begin{align}
    \hat{A}_{iso}=\frac{8\pi}{3}A_0 \sum_{\beta,\alpha=s}|\beta\rangle\langle \beta|\delta(\mathbf{r}-\mathbf{R}_q)\label{eq:A_iso}
\end{align}
and
\begin{align}
    \hat{A}_{ij} = A_0 \frac{\hat{Q}^{(2)}_{ij}}{r^3},\label{eq:A_aniso}
\end{align}
respectively, where $\hat{Q}^{(2)}_{ij}=3\hat{r}_i\hat{r}_j-\delta_{ij}$ is the quadrupole moment tensor, and $A_0=g_s g_N \mu_0 \mu_B \mu_N/4\pi \langle S_z\rangle$. The sum in the isotropic hyperfine tensor equation selectively projects out contributions from $|s\rangle$ orbitals.

Evaluating  Eqs. (\ref{eq:A_iso}-\ref{eq:A_aniso}) in the LCAO basis yields,
\begin{align}
    \langle\hat{A}_{iso}\rangle_\omega=\frac{-8}{3}A_0 \sum_{q}Im[Tr[G_{s,s}(\mathbf{R}_q,\mathbf{R}_q,\omega)]]
\end{align}
\begin{align}
    \langle A_{ij}\rangle_\omega= \frac{-1}{\pi} Im\left[\sum_{\beta,\beta'}A^\beta_0  c^*_{\beta'}c_{\beta} \langle \beta'|\frac{\hat{Q}^{(2)}_{ij}}{r^3}|\beta\rangle\right].\label{eq:TB_hyperfine}
\end{align}
where $c_{\beta}$ are the tight-binding coefficients ($|\psi\rangle=\sum_{\beta} c_{\beta} |\beta\rangle$, and we recognize $c^*_{\beta'}c_{\beta}$ as the inhomogeneous Green's function solution, $G_{\beta',\beta}$. The second term can be decomposed as the product of a quadrupole moment tensor and a physical parameter corresponding to the size of the atomic orbital,

\begin{figure}[h!]
    \centering
    \includegraphics[width=.9\linewidth]{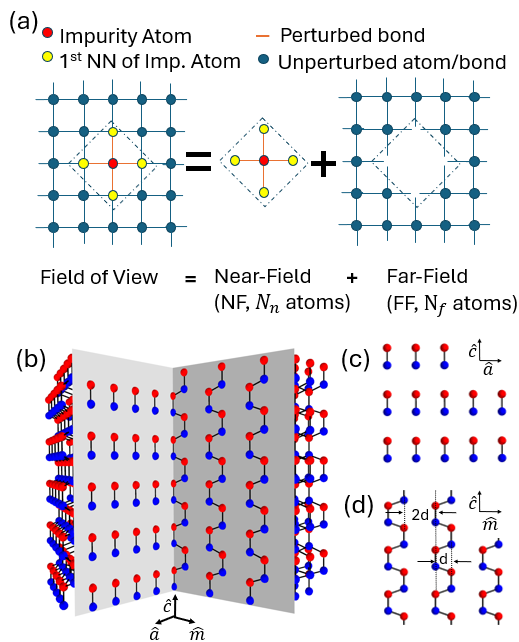}
    \caption{Schematic depicting (a) the separation of the lattice sites into the near field and far field based on the region of space containing the perturbation, (b) a 3D representation of the $\hat{c}$, $\hat{m}$ and $\hat{a}$ planes, and planes of constant height defined by (c) an $\hat{m}$ cleave and (d) an $\hat{a}$ cleave.  Figure (d) shows the nature of the alternating plane height observed for slices along $\hat{m}$ ($d=0.92059$ \AA).}
    \label{fig:near_far_cartoon}
\end{figure}

\begin{align}
\langle \beta'|\frac{\hat{Q}^{(2)}_{ij}}{r^3}|\beta\rangle=\langle r^{-3}\rangle_{\beta',\beta}\langle \hat{Q}_{ij}\rangle_{\beta',\beta}
\end{align}
where
\begin{align}
\langle \hat{Q}_{ij}\rangle_{\beta',\beta}=\int \langle l'\alpha'|\hat{Q}_{ij}|l\alpha \rangle d\Omega
\end{align}
 and 
\begin{align}
\langle r^{-3}\rangle_{\beta',\beta}=\int \langle R_{n'l'}|r^{-3}|R_{nl} \rangle r^2dr.
\end{align}
The $Q_{ij}$ matrices are tabulated in the Supplemental\cite{SuppFootnote}. For the $\langle r^{-3}\rangle$ values, we use the atomic values from Ref. \cite{KOHMiller.sciencedirect.1985}.
The anisotropic hyperfine tensor can then be written as,
\begin{align}
    \langle A_{ij}\rangle_\omega= \frac{-1}{\pi} Im\left[\sum_{\beta,\beta'} A^\beta_0 G_{\beta',\beta}(\omega) \langle r^{-3}\rangle_{\beta',\beta} \langle \hat{Q}_{ij}\rangle_{\beta'\beta}\right]\label{eq:TB_hyperfine_spherical} 
\end{align}

The term $\langle r^{-3}\rangle_{\beta',\beta}$ between atoms located at different atomic sites (i.e., $q\neq q'$) is expected to be negligibly small. 
Only contributions satisfying $q=q'$ are considered in the present work.

\section{Results and Discussion}\label{sec:results}

\begin{table}[h!]
    \centering
    \begin{tabular}{c c c c}
    [I,J,K,L]&$r_0$ (\AA)&$A_{qq}$ (MHz)&$[u_x,u_y,u_z]_q$\\
    \hline
    \hline
\multicolumn{4}{l}{\bf 1NN Axial - Nitrogen}\\
~$[ 0, 0, 0, 0]$&~~1.9540~~&~~~-6.4265~~~&[ 0.8610,~ 0.5086,~ 0.0000]\\
             &&-6.4265&[ 0.5086,~-0.8610,~ 0.0000]\\
             && 23.7987&[-0.0000,~-0.0000,~ 1.0000]\\
\multicolumn{4}{l}{\bf 1NN Basal - Nitrogen}\\
~$[ 0,  0,  0,  2]$&1.9487&-0.7025&[-1.0000,~-0.0002,~-0.0005]\\ 
             &&-0.7020&[-0.0006,~ 0.2956,~ 0.9553]\\ 
             && 2.6639&[ 0.0000,~-0.9553,~ 0.2956]\\
~$[\overline{1}, 0, 0, 2]$&1.9487&-0.7012&[-0.5000,~ 0.8660,~ 0.0002]\\ 
             &&-0.7007&[-0.2559,~-0.1479,~ 0.9553]\\ 
             && 2.6591&[-0.8273,~-0.4777,~-0.2956]\\ 
~$[ 0,\overline{1}, 0, 2]$&1.9487&-0.7025&[ 0.5000,~ 0.8660,~-0.0002]\\ 
             &&-0.7020&[ 0.2561,~-0.1477,~ 0.9553]\\ 
             && 2.6639&[-0.8273,~ 0.4777,~ 0.2956]\\

\multicolumn{4}{l}{\bf 2NN - Gallium}\\
~$[\overline{1}, 0,\overline{1}, 3]$&3.1798&-0.1628&[-0.6076,~ 0.6471,~-0.4606]\\
             && 0.8661&[ 0.4645,~ 0.7599,~ 0.4547]\\
             && 2.4506&[-0.6442,~-0.0624,~ 0.7623]\\
~$[ 0,\overline{1},\overline{1}, 3]$&3.1798&-0.1625&[ 0.6076,~ 0.6470,~-0.4605]\\
             && 0.8662&[-0.4645,~ 0.7599,~ 0.4547]\\
             && 2.4509&[-0.6442,~ 0.0624,~-0.7623]\\
~$[ 0, 0,\overline{1}, 3]$&3.1798& 0.0905&[ 1.0000,~-0.0000,~-0.0000]\\
             && 1.0681&[ 0.0000,~ 0.7346,~ 0.6784]\\
             && 1.9960&[-0.0000,~-0.6784,~ 0.7346]\\

 ~$[ 0, 0, 0, 3]$&3.1798&-0.1129&[-1.0000,~ 0.0000,~ 0.0000]\\
                &&-0.0217&[-0.0000,~-0.8482,~ 0.5296]\\
               && 0.1743&[-0.0000,~-0.5296,~-0.8482]\\
~$[\overline{1}, 0, 0, 3]$&3.1798&-0.1078&[ 0.4174,~ 0.8917,~ 0.1751]\\
             && 0.0180&[ 0.9056,~-0.3923,~-0.1611]\\
             && 0.1293&[ 0.0750,~-0.2258,~ 0.9713]\\
~$[ 0,\overline{1}, 0, 3]$&3.1798&-0.1080&[-0.4176,~ 0.8916,~ 0.1749]\\
             && 0.0181&[ 0.9055,~ 0.3925,~ 0.1614]\\
             && 0.1296&[-0.0752,~-0.2258,~ 0.9713]
    \end{tabular}
    \caption{Anisotropic hole hyperfine ($A_{ii}$) and unit vectors ($\hat{u}_i$) for $V_{Ga}$ out to second-nearest neighbors. $[I,J,K,L]$ labels a unique atomic site ($\mathbf{r}=I\mathbf{a}_1+J\mathbf{a}_2+K\mathbf{a}_3+\mathbf{t}_L$). }
    \label{tab:Ga_vac_table}
\end{table}

~

\begin{table}[h!]
    \centering
    \begin{tabular}{c c c}
    ~[I,J,K,L]&$r_0$ (\AA)&$A_{iso}$ (MHz)\\
    \hline
    \hline
\multicolumn{2}{l}{\bf 1NN Axial   - Gallium }\\
~$[ 0, 0, 0, 1]	$	&1.9540&80.6125\\
\multicolumn{2}{l}{\bf 1NN Basal - Gallium   }\\
~$[ 0,\overline{1},\overline{1}, 3],[\overline{1}, 0,\overline{1}, 3],[ 0, 0,\overline{1}, 3]$	&1.9487&84.8936\\

\multicolumn{2}{l}{\bf 2NN  -  Nitrogen}\\
~$[\overline{1}, 0, 0, 2],[ 0, 0,\overline{1}, 2],[ 0,\overline{1}, 0, 2]$	&3.1798&	2.1615\\
~$[\overline{1}, 0,\overline{1}, 2],[ 0, 0, 0, 2],[ 0,\overline{1},\overline{1}, 2]$	&\\
\\
~$[ 0,\overline{1}, 0, 0],[\overline{1}, 0, 0, 0],[\overline{1}, 1, 0, 0]$	&3.1890&	2.1684\\
~$[ 0, 1, 0, 0],[ 1, 0, 0, 0],[ 1,\overline{1}, 0, 0]$	&\\

\multicolumn{2}{l}{\bf N$^{th}$-NN Gallium  Shells}\\
~$[ 0, 0,\overline{1}, 1]$	&3.231&	0.6585\\
\\
~$[\overline{1}, 1,\overline{1}, 3],[ 1,\overline{1},\overline{1}, 3],[\overline{1},\overline{1},\overline{1}, 3]$	&3.7373&	1.8286\\
\\
~$[ 0,\overline{1}, 0, 1],[\overline{1}, 0, 0, 1],[\overline{1}, 1, 0, 1]$	&3.7400	&2.3036\\
~$[ 0, 1, 0, 1],[ 1, 0, 0, 1],[ 1,\overline{1}, 0, 1]$		&\\
\\
~$[ 0,\overline{1},\overline{1}, 1],[\overline{1}, 0,\overline{1}, 1],[ 0, 1,\overline{1}, 1]$	&4.5397&3.6904\\
\\
~$[ 1, 0,\overline{1}, 1],[\overline{1}, 1,\overline{1}, 1],[ 1,\overline{1},\overline{1}, 1]$&4.5397&	3.6853\\
\\
~$[ 0,\overline{1}, 0, 3],[ 0, 0, 0, 3],[\overline{1}, 0, 0, 3]$	&4.9052&	5.065\\
\\
~$[ 0, 1,\overline{1}, 3],[\overline{2}, 1,\overline{1}, 3],[ 1, 0,\overline{1}, 3]$	&4.9129&	5.8326\\
~$[ 0,\overline{2},\overline{1}, 3],[\overline{2}, 0,\overline{1}, 3],[ 1,\overline{2},\overline{1}, 3]$		&\\
\\
~$[\overline{1}, 1, 0, 3],[ 1,\overline{1}, 0, 3],[\overline{1},\overline{1}, 0, 3]$	& 5.8507&0.974\\
\\
~$[\overline{2}, 1, 0, 1],[ 1,\overline{2}, 0, 1],[ 1, 1, 0, 1]$	&~~5.8589~~&	0.9941\\
~$[ 2,\overline{1}, 0, 1],[\overline{1}, 2, 0, 1],[\overline{1},\overline{1}, 0, 1]$&
    \end{tabular}
    \caption{Isotropic electron hyperfine for the $V_{N}$ defect in excess of 0.5 MHz. $[I,J,K,L]$ labels a unique atomic site ($\mathbf{r}=I\mathbf{a}_1+J\mathbf{a}_2+K\mathbf{a}_3+\mathbf{t}_L$). A complete table of couplings for the calculation can be found in the Supplemental Material.}
    \label{tab:N_vac_table}
\end{table}

\subsection{Nitrogen and Gallium Vacancies}

The gallium and nitrogen vacancies, $V_{Ga}$ and $V_{N}$,  are expected to produce a deep acceptor around 225 meV\cite{Monemar.JAP.1980} and a donor in the range of 130 to 30 meV\cite{Buckeridge.PRL.2015,Look.APL.2003,Horita.APL.2021}, respectively. Both bound states energies correspond with the hydrogenic defect model limit, 
\begin{equation}
    E_d=\frac{m^*R_\infty}{m_0 \epsilon^2_{\omega\rightarrow 0}}. 
\end{equation}

We model these vacancies using the potential,
\begin{align}
\hat{V}'=\delta E-\sum_{\langle NN\rangle}V^{Bulk}_{\langle NN\rangle}
\end{align}
where $-V^{Bulk}_{\langle NN\rangle}$ corresponds to the negative of the GaN electronic TB parameters from Ref. \onlinecite{Jancu.APL.2002} that couple the vacancy site to its nearest neighbors. This potential can be understood intuitively as decoupling the four nearest neighbor Ga-N bonds from the host site, resulting in a set of atomic states electronically isolated from the rest of the system in addition to 4 dangling bonds at the nearest neighbors. A large onsite shift $\delta E$ is then introduced to place these unphy              sical states far from the Fermi level, effectively projecting them out of the Hilbert space of the crystal.

For the ideal $V_{Ga}$ defect, we calculate a hole binding energy for the $V_{Ga}$ of 238 meV, in agreement with experiment and the A-band (heavy hole) hydrogenic acceptor binding energy 240 meV ($m_A^*=1.4, \epsilon_{\omega\rightarrow 0}=8.9$\cite{levinshtein2001properties}).

The $V_{N}$ defect nearest neighbor positions have been hydrostatically expanded around the vacancy site by $1.25\%$ resulting in an electron binding energy of 31 meV. This trend is in agreement with {\it ab initio} calculations, which found similar values for the Ga bond hydrostatic strain for $V_N$\cite{limpijumnong2004diffusivity}. We calculate the deformation potential using the strain parameters in Ref. \cite{Jancu.APL.2002}.
The theoretical conduction band hydrogenic ideal donor binding energy is 34 meV ($m_e^*=0.2, \epsilon_{\omega\rightarrow 0}=8.9$\cite{levinshtein2001properties}).

The wave functions for both defects are depicted using Eq. (\ref{eq:LDOS}) in Fig. \ref{fig:wavefunction}. The $V_{Ga}$ wave function is highly anisotropic with predominantly p-orbital character, whereas the $V_{N}$ wave function is largely isotropic and significantly more delocalized with s-orbital character.

Using Eq. (\ref{eq:TB_hyperfine_spherical}), we find the hyperfine tensor components for the first nearest neighbor shell of the gallium vacancy to be $\{A_{\perp},A_{\parallel}\}=\{-0.555,1.920\}$ MHz for the basal plane neighbors and $\{A_{\perp},A_{\parallel}\}=\{-4.427,14.99\}$ MHz for the axial bonded neighbor. For the shallow nitrogen vacancy, we find the basal and axial isotropic hyperfine couplings to be $84.89$ MHz and $80.61$ MHz, respectively.

\section{Conclusion}
An accurate description of the local hyperfine coupling of a defect to the nuclear bath is necessary for distinguishing between defect states. To this end, we present calculations of the near field wave functions and hyperfine couplings for the $V_{Ga}$ and $V_{N}$ native point defects in GaN utilizing a real space spds$^*$ with spin orbit tight binding  Green's functions formalism. We have found that the nitrogen vacancy to be a shallow donor with a highly isotropic wave function and near neighbor isotropic hyperfine basal and axial isotropic hyperfine couplings to be $84.89$ MHz and $80.61$ MHz, respectively.
Conversely, the $V_{Ga}$ defect was found to be a deep acceptor with a highly anisotropic wave function and first nearest neighbor $\{A_{\perp},A_{\parallel}\}=\{-0.555,1.920\}$ MHz for the basal plane neighbors and $\{A_{\perp},A_{\parallel}\}=\{-4.427,14.99\}$ MHz for the axial bonded neighbor.

\section{Acknowledgments}
This material is based upon work supported by the Air Force Office of Scientific Research under award number FA9550-22-1-0308.

\section{Data Availability}
The data that support the findings of this article are openly available at https://doi.org/10.5281/zenodo.19263597.

\providecommand{\noopsort}[1]{}\providecommand{\singleletter}[1]{#1}

\end{document}